\documentclass[aps,prl,twocolumn,showpacs]{revtex4}
\usepackage{amsmath}
\usepackage{graphicx,epsfig,psfrag}
\usepackage{amssymb}

\renewcommand{\vec}[1]{{\bf #1}}
\begin{document}

\title{Mott transition of fermionic atoms in a three-dimensional
  optical trap}
\author{R. W. Helmes${}^1$}
\author{T. A. Costi${}^2$}
\author{A. Rosch${}^1$}
\affiliation{\hspace*{-1.9cm}${}^1$Institute for Theoretical Physics, University of Cologne, 50937
Cologne, Germany\hspace*{-1.9cm} \\
${}^2$Institute of Solid State Research, Research Centre J\"ulich, 52425 J\"ulich, Germany}
\begin{abstract}
  We study theoretically the Mott metal-insulator transition for a
  system of fermionic atoms confined in a three-dimensional optical
  lattice and a harmonic trap. We describe an inhomogeneous system of
  several thousand sites using an adaptation of dynamical mean field
  theory 
  solved efficiently with the numerical renormalization group method.
  Above a critical value of the on-site interaction, a Mott-insulating
  phase appears in the system. We investigate signatures of the Mott
  phase in the density profile and in time-of-flight experiments.
\end{abstract}

\date{\today}
\maketitle

{\em Introduction:}
Ultracold atoms in optical lattices offer exciting possibilities to investigate many-particle effects
and to
realize and measure models of condensed matter physics like the Hubbard model
with unprecedented control of the band structure and  interaction strength.

One of the most dramatic effects of strong correlations is the Mott
transition, where strong interactions drive a system insulating.
While for bosonic atoms \cite{JakschPRL1998} the Mott transition in an
optical lattice has been realized a few years ago by Greiner {\it et
  al.} \cite{GreinerNature2002}, the corresponding experiment for
fermionic systems turns out to be much more difficult, partially due
to problems with cooling and due to the need to work with two
fermionic species to model the spin degree of freedom.  However, due
to the enormous experimental progress, a realization of the Mott
transition in fermionic systems is expected in the near future. For
example, recently K\"ohl {\it et al.} \cite{KohlPRL2005} succeeded to
capture a single species of fermions in an optical lattice and to
observe the Fermi surface. Also two species of fermions have been
successfully trapped and cooled down in a series of experiments
studying e.g. the BEC-BCS crossover \cite{Regal2004}.

For the interpretation of the experiments it is essential to
investigate the effects of the smooth external confining potential holding the atoms in the trap.  The resulting inhomogeneities can
make it more difficult to interpret the experiments, but
induce also new interesting effects, e.g. associated to the sharp
surface between metallic and insulating regions.  A frequently used
approximation (e.g. in the present context by Ref.~\cite{LiuPRL2005})
is to describe the trapped atoms locally by a homogeneous system as in
the local density approximation.  However, in the presence of
sharp domain walls between two phases, such an approximation is
expected to fail and a more realistic treatment of the inhomogeneous
system is necessary.

For one-dimensional (1D) systems, powerful numerical \cite{RigolPRL2003,
  RigolPRB2006} and analytical \cite{LiuPRL2005} methods exists to
study theoretically the Mott transition in fermionic systems. For
example, in Refs.~\cite{RigolPRL2003, RigolPRB2006,LiuPRL2005} quantum
Monte Carlo techniques were used to investigate the signatures of Mott
phases in the presence of an external harmonic confinement potential
for a 1D system. Rigol {\it
  et al.}  \cite{RigolPRL2003, RigolPRB2006} argued, that in one
dimension the inhomogeneities resulting from the trapping potential
essentially destroy the main signatures of Mott phases in
time-of-flight experiments.

As exact numerical methods for fermions can only be applied to very
small systems, one has to resort to approximations to calculate the
properties of three dimensional lattices of realistic size. Here the
method of choice is the so-called dynamical mean field theory (DMFT)
\cite{MetznerVollhardt,GeorgesRevMod1996}. Within DMFT, the only
approximation is to neglect non-local contributions to the
self-energy. This allows to map the $N$ site lattice problems to $N$
single-impurity Anderson models coupled by a self-consistency
condition, see discussion below. DMFT is, for example, frequently used
to describe complex bulk materials, e.g. by combining DMFT with
band-structure calculations to obtain an ab-inito description of
strongly correlated materials.  In a few cases, DMFT has been employed
to describe inhomogeneous
systems \cite{Dobrosavljevic1997,Potthoff,Freericks,Okamoto2004}
like the surface of Mott insulators \cite{Okamoto2004} or disordered
 materials \cite{Dobrosavljevic1998}.

 A main problem of DMFT is the need for an reliable and efficient
 method to solve the effective impurity problems. Previous
 applications of DMFT to inhomogeneous systems were using impurity
 solvers like a two-site approximation \cite{Okamoto2004} or slave-boson
 mean-field theory \cite{Dobrosavljevic1998}, implying severe
 further approximations, or started from simplified fermionic models
 such as the Falicov-Kimball model \cite{Freericks}. We will show
 that one can also use
 efficiently one of the most accurate impurity solvers, the numerical
 renormalization group \cite{BullaPRB2001,Bulla2007} (NRG), to obtain
 reliable results for traps containing several thousand atoms 
modeling a fermionic Hubbard model.

 After introducing the model and our method (DMFT for
 inhomogeneous systems+NRG), we will show the resulting spectral
 functions and discuss how the transition from a metal to a Mott
 insulating phase can be seen in real-space and time-of-flight
 experiments.
We investigate the role of temperature, filling and interaction strength.

{\em Model and Method:}
We consider the fermionic Hubbard model on a 3D-cubic lattice,
\begin{align}
  {\cal{H}} =- J \sum_{\langle ij \rangle,\sigma} c_{i\sigma}^{\dag} c_{j\sigma} +
  U \sum_i n_{i\uparrow} n_{i\downarrow} +
 V_0  \sum_{i,\sigma} r_i^2  n_{i\sigma},\label{h}
\end{align}
where $c_{i\sigma}^{\dag}$ creates a fermion at site $i$ with spin
$\sigma$, $n_{i\sigma} = c_{i\sigma}^{\dag} c_{i\sigma}$ is the local density, $J$ the
 nearest-neighbor tunneling matrix element and $U$  the effective
 on-site interaction and the lattice distance is set to unity.
 We include  4224 sites with a distance $r_i\le R=10$ from the center
 of the trap, located in the middle of 8 central sites. The
 strength of the confining harmonic potential $V_0=0.276\,J$ is chosen such
 that all sites with $r_i>R$ are unoccupied and can be neglected.
 Using the symmetry of the cubic
 lattice, one has to deal with only $118$ inequivalent sites.

\begin{figure}
\includegraphics[width=\linewidth,clip]{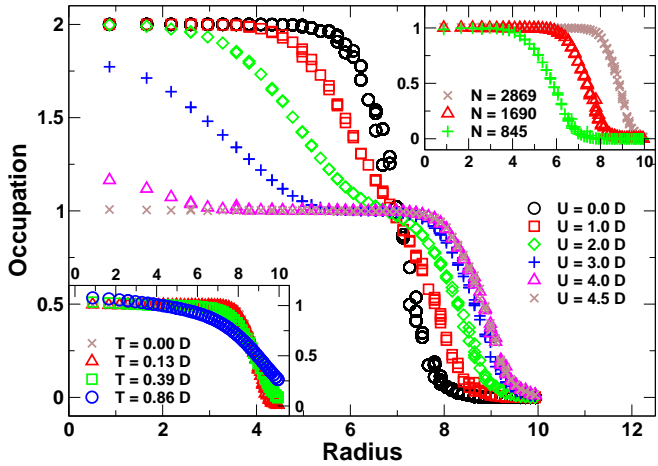}
\caption{{\it (color online)} The number of fermions $\langle n_i
  \rangle$ per lattice site, as a function
  of  the distance $r$ to the origin. The signature of the Mott
  insulating phase is the presence of a plateau with $\langle n_i
  \rangle=1$. Note that there are inequivalent
  sites with different occupation but the same $r$. Main panel: Crossover from weak to
  strong interactions for a fixed number of $N=2869$ particles in the
  trap at $T=0$, $U=6 J$.
 Upper inset: Dependence on the number of particles for $U=4.5 D, T=0$. Lower inset: $T$            dependence for $U=4.5 D$, $N=2869$ (the $T=0$ and
 $T=0.13 D$ curves lie on top of each other).}
\label{Occupation}
\end{figure}
 The basic idea of DMFT  \cite{GeorgesRevMod1996} is to pick 
 a single site of the lattice (the 'impurity' $i$) and model the effect
 of all the other sites by a {\em non-interacting} bath of fermions.
 The resulting Anderson impurity model is defined by the local
 interaction $U$ and the hybridization of the impurity to the bath.
 The latter is encoded in the $U=0$ Green's function,
 $G^0_{{\rm And},i}(\omega)$. For this model, one determines the local
 self-energy $\Sigma_i(\omega)$ (see below). From the
 $\Sigma_i(\omega)$, one can construct the lattice Green's function
\begin{equation}
  {{G_{\rm{lat}}}^{-1}}(\omega)_{ij} = \delta_{i,j}\left( \omega + \mu -
  \Sigma_i(\omega)-V_0 r_i^2 \right) - J_{ij}, \label{lat}
\end{equation}
where $J_{ij}=J$ if sites $i$ and $j$ are nearest neighbors and $0$
otherwise.  The bath of each impurity is then determined from the
requirement that at each site the lattice Green's function and the
Green's function of the impurity model coincide,
$G_{\rm{lat}}(\omega)_{ii} =\left[ {{\cal{G}}^0_{{\rm
        And},i}}(\omega)^{-1} - \Sigma_i(\omega)\right]^{-1} $, thus
establishing a self-consistency loop.  The scheme described above scheme can
be derived using as the {\em only} approximation that the self-energy
is a local quantity.  Both the non-local single particle quantum
mechanics  of fermions and all local effects of
strong interactions are correctly described within DMFT.

A main difficulty of DMFT is, however, an accurate calculation of the
self-energy of the Anderson impurity model. For this we use the
NRG  
\footnote{We
  use for all of our calculations (see Ref.~\cite{Bulla2007}) $\Lambda=1.5$, 60 shells and 800
  kept states (using particle number, $S_z$ and $S^2$ conservation).},
  see Ref.~\cite{Bulla2007} for a description of the method. To
  obtain efficiently $G_{\rm lat}$ from an inversion of Eq.~(\ref{lat}) it is
  essential to use the full symmetry of the cubic lattice.

In this paper we restrict ourselves to paramagnetic solutions which
simplifies the rather challenging numerics considerably. Also
experimentally, it is very difficult to reach the low temperatures
below which magnetism is expected.
Furthermore, we do not expect that magnetic order will change the
density profiles or time-of-flight pictures considerably.

\begin{figure}
\includegraphics[width=\linewidth]{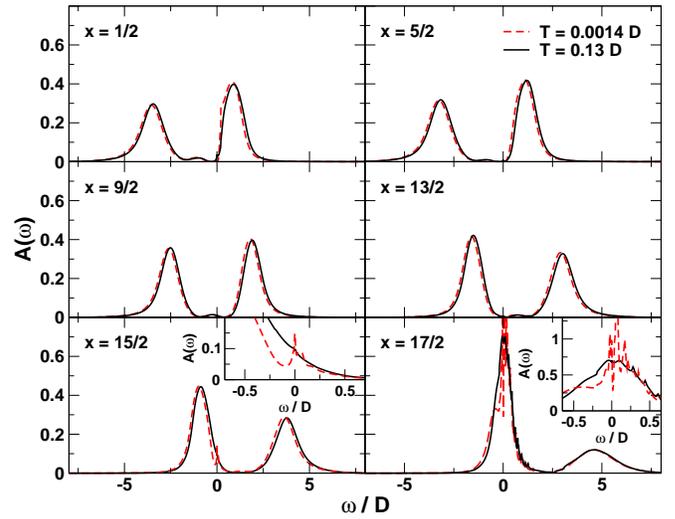}
\caption{{\it (color online)} Local spectral functions in the Mott insulating phase ($U=4.5
  D, N=2869$) for two different $T$ for lattice points 
with coordinates $(x,1/2,1/2)$. Left inset: The coherence peak at the
Fermi energy,
characteristic for a strongly correlated metal, vanishes with
increasing $T$. Right inset: Close to edge of the atomic cloud, where
the potential becomes steep, finite
size effects are visible.}
\label{SpectralFunctions}
\end{figure}

{\em Results:} 
Fig.~\ref{Occupation} shows how the number of fermions per site
evolves for increasing interactions $U/J$, which push the fermions away from the center of the
trap.  For the chosen parameters, we obtain for $U/D=0,1,2$ ($D=6 J$
is half the bandwidth) a band insulator in the center of the trap and a
metal further outside. For the homogeneous system the critical
interaction is given by $U_c/D=2.52$, but already for $U/D=2$ the
compressibility close to half-filling is strongly reduced as can be
seen in a shoulder in the curve for $\langle n_i \rangle\approx 1$.
For $U/D\ge 3$ the incompressibility of the Mott insulating state, $\partial n/\partial \mu=0$,
manifests itself in a plateau. The thickness of this Mott insulating 'onion
shell' increases for increasing $U$ eliminating the metallic phase in
the center for $U/D=4.5$. The insets of
Fig.~\ref{Occupation} show how the Mott insulating region at $U/D=4.5$
shrinks again when the number of fermions in the trap is reduced and
how thermal excitations destroy the Mott plateau.

Our method allows to study the spatial dependence of the spectral functions. While this quantity
is difficult to measure for atoms in a trap, it is a highly sensitive
probe of the metal-insulator
transition. Fig~\ref{SpectralFunctions} shows how the local spectral
function evolves when moving from the center to the edge of the trap
at $U/D=4.5$. In the insulating regime, the spectral function
$A(\omega)$ is
characterized by the two Hubbard bands with equal weight and 
$A(\omega=0)$ becomes very small (it never vanishes exactly as atoms from
the metallic regions can tunnel into the insulator). As a function of
the distance from the center, the Hubbard bands
shift due to the harmonic potential.
When the Fermi energy starts to merge with one of the Hubbard bands,  
a sharp quasi-particle peak emerges at $\omega=0$
for sufficiently low $T$ (see inset). 

\begin{figure}
\includegraphics[width=\linewidth]{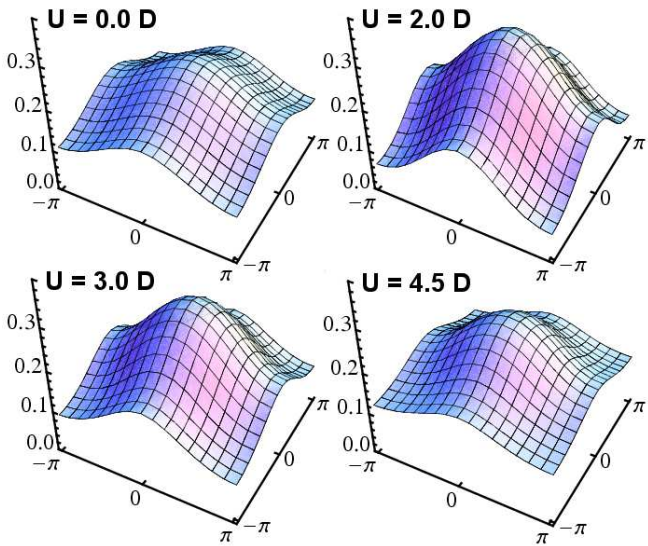}
\caption{{\it (color online)} Momentum distribution $n^{\rm tof}_{\vec{k}}$ for
  $U/D=0,2,3,4.5$. Both in the predominantly band-insulating phase
  ($U/D=0$) and Mott insulating phase ($U/D=4.5$) the curves are considerably
  flatter than for $U/D=2,3$ where most fermions are in the metallic phase. }
\label{TOF}
\end{figure}
In a time-of-flight experiment, the two-dimensional projection, 
$n^{\rm tof}_{\vec{k}}=\int n_\vec{k} d
k_z/(2 \pi)$, of the three-dimensional momentum distribution 
\begin{eqnarray}
n_\vec{k}&=& -\frac{1}{N} \sum_{i,j} \int \frac{d \omega}{\pi} 
f(\omega) e^{i \vec{k} (\vec{r}_i-\vec{r}_j)} \text{Im} G_{ij}(\omega)
\end{eqnarray}
of the fermions can be measured \cite{GreinerNature2002,Gerbier2007}. Here $f(\omega)$
is the Fermi function and we have normalized $n_\vec{k}$ such that
$\int n_\vec{k} \frac{d^3 \vec{k}}{(2 \pi)^3}=\int n^{\rm tof}_\vec{k}
\frac{d^2 \vec{k}}{(2 \pi)^2}=1$.

\begin{figure}
\includegraphics[width=\linewidth]{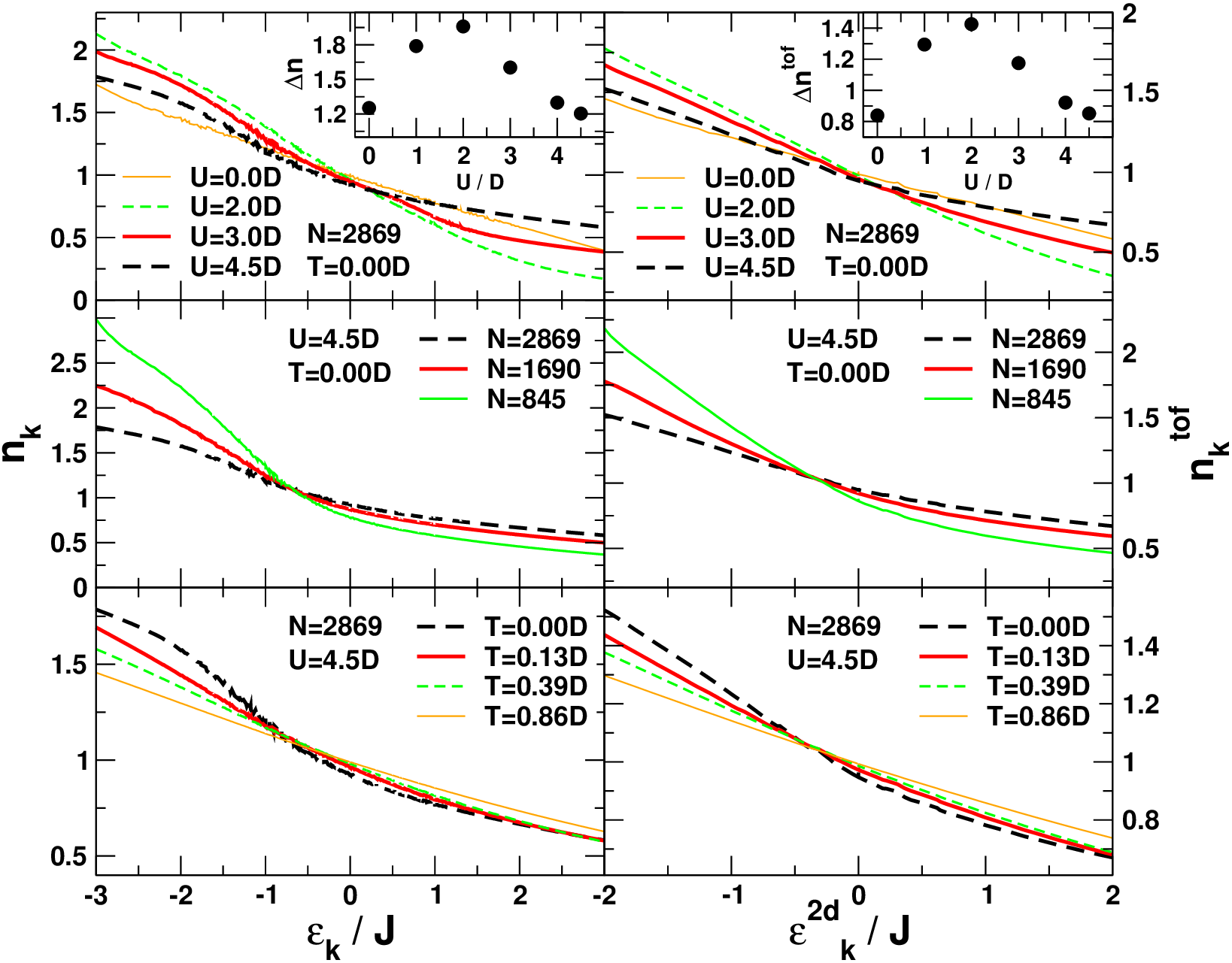}
\caption{{\it (color online)} $n_{\vec{k}}$ (left panel) and $n^{\rm tof}_{\vec{k}}$ (right
  panel) for $\vec{k}=\pi/10 (n_x,n_y,n_z)$, $n_{x,y,z}=-10,...,10$,
  plotted as a function of $\epsilon_\vec{k}=-2 J (\cos k_x +\cos k_y
  +\cos k_z)$ and $\epsilon_\vec{k}^{2d}=-2 J (\cos k_x +\cos k_y)$,
  respectively, for different values of $U$, $N$ and $T$ (upper,
  middle and lower panels, respectively). For a given value of
  $\epsilon_\vec{k}$ a range of values of $n(\epsilon_\vec{k})$
  exists. Nevertheless, each curve collapses to a single line in good
approximation. Inset:  $\Delta n=n_{0,0,0}-n_{\pi,\pi,\pi}$ (and
$\Delta n^{\rm tof}=n^{\rm tof}_{0,0}-n^{\rm tof}_{\pi,\pi}$) as a function of
$U$. $\Delta n$ and $\Delta n^{\rm tof}$ are largest for the predominantly 
metallic phases and smallest  for phases with a large band-
(small $U$)  or Mott-insulating (large $U$) regions.}
\label{TOF2}
\end{figure}

Fig.~\ref{TOF} shows $n^{\rm tof}_{\vec{k}}$. For a quantitative
analysis of the results and for comparison to (future) experiments, we
suggest to plot $n^{\rm tof}_{\vec{k}}$ as a function of the
tight-binding dispersion $\epsilon^{2d}_\vec{k}=-2 J (\cos k_x +\cos
k_y)$, see Fig.~\ref{TOF2}. Surprisingly, each curve collapses to good
approximation to a single line, despite the fact that for a given
$\epsilon^{2d}_\vec{k}$ a range of $n^{\rm tof}_\vec{k}$ exists (as is noticeable
    in the small scatter of the curves for $n_{\vec{k}}$). Here it is
useful to remember that within DMFT, $n_{\vec{k}}$ is only a function
of $\epsilon_\vec{k}$ for the {\em homogeneous} system. Therefore, if a
local density approximation (LDA) were exactly valid, the collapse to
a single curve would be perfect. Therefore our results suggest, that LDA
is a very good approximation for the analysis of time-of-flight pictures (but not for other
quantities, see below).

It is an interesting but difficult question whether this effect is
partially an artifact of DMFT which neglects the momentum dependence
of the self-energy. While for the experimentally relevant temperature
range this is probably a very good approximation, it is expected to
fail very close to the metal-insulator transitions at low $T$.

Qualitatively, the results of Fig.~\ref{TOF2} reflect that localized
electrons in the band- or Mott insulating phase are characterized by a
momentum-independent $n_\vec{k}$ while in the homogeneous metallic
phase $n_\vec{k}$ displays a jump at the Fermi momentum with a height
given by the quasi-particle weight $Z$. As the effective local Fermi
momentum varies smoothly within the trap, all jumps are smeared out.
However, the slope of $n(\epsilon_\vec{k})$ or the difference $\Delta
n^{\rm tof}=n^{\rm tof}_{0,0}-n^{\rm tof}_{\pi,\pi}$ is still a good
measure of how metallic or insulating the system is. The inset of
Fig.~\ref{TOF2} describes the evolution from a mainly band-insulating
via a dominately metallic to a Mott insulating regime when $U/D$ is
increased. Similarly, the middle panel of Fig.~\ref{TOF2} shows how
$\Delta n^{\rm tof}$ increases when at large $U/D$ the number of
particles and therefore the size of the Mott insulating region is
reduced (compare with upper inset of Fig.~\ref{Occupation}). For
increasing temperature (lower panel of Fig.~\ref{TOF2}) the
destruction of quantum coherence leads to a flattening of
$n^{\rm tof}_\vec{k}$. Note that  $n^{\rm
  tof}_\vec{k}$ is more sensitive to changes of $T$ compared to
$\langle n_i \rangle$, see Fig.~~\ref{Occupation}.

\begin{figure}
\includegraphics[width=0.9 \linewidth,clip]{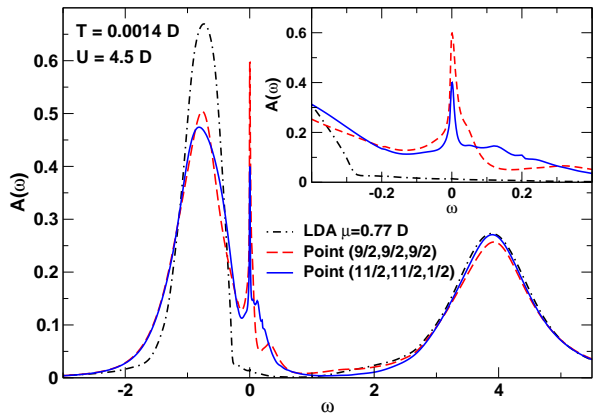}
\caption{{\it (color online)} Local spectral function of two non-equivalent lattice points
  at distance $7.794$ from the origin at the boundary of the Mott
  insulating region ($T=0.0014 D$, $U = 4.5 D$). Here the LDA
  (dot-dashed line) fails
  completely to describe the spectral function but still reproduces
  the occupation with high accuracy
  ($n_{1/2(9,9,9)}=0.481,n_{1/2(11,11,1)}=0.482,n_{\rm LDA}=0.496$).}
\label{LDAComparison}
\end{figure}
{\em Conclusions:} In this paper we investigated the signatures of the
Mott transition of fermions in an optical trap using a local
approximation to the self energy (space-resolved DMFT+NRG) which
allows to treat several thousand atoms. The clearest signature of a
Mott phase is a plateau in the density profile $n(\vec{r})$ of the atoms, see
Fig.~\ref{Occupation}. These plateaus are, however, washed out if only the
column density, $\int dz \hspace{.05cm} n(\vec{r})$, which can be measured directly, is considered (not shown). 
The Mott transition is more difficult to observe in a time-of-flight
experiment. However, the insets of Fig.~\ref{TOF2} show that a
characteristic flattening of $n^{\rm tof}_\vec{k}$ can be seen when
a large fraction of the trap becomes a band or Mott insulator. 

Our calculations did not rely on a local density approximation (LDA)
which allows us to investigate whether this widely used approximation
is valid in the present context. It turns out that both density
profiles and TOF experiments are rather well described by LDA. As
discussed in the introduction, LDA is expected not to be valid close
to a sharp domain boundary.  Indeed, Fig.~\ref{LDAComparison} shows
that the LDA fails completely to describe the low-energy excitation
spectrum at the boundary of the Mott insulating region. The coherence
peak at the Fermi energy arises due to the penetration of the metallic
phase into the Mott insulator via the Kondo effect. 

For the future it will be interesting to investigate the effects of
magnetism. In systems with a population imbalance we expect that the
majority spin will accumulate in the Mott insulating regions. These
effects will be studied in a forthcoming publication.  

We acknowledge useful discussions with L.~Craco, T.~Micklitz, D.~van
Oosten, H.~Tjeng, M.~Vojta, supercomputer support by the John von
Neumann institute for Computing (J\"ulich) and the Regional Computing
Center Cologne and financial support by the SFB 608 of the DFG.


\begin{thebibliography}{10}

\bibitem{JakschPRL1998}
D. {Jaksch} {\it et~al.}, Phys. Rev. Lett. {\bf 81},  3108  (1998).

\bibitem{GreinerNature2002}
M. {Greiner} {\it et~al.}, \nat {\bf 415},  39  (2002).

\bibitem{KohlPRL2005}
M. {K{\"o}hl} {\it et~al.}, Phys. Rev. Lett. {\bf 94},  080403  (2005).

\bibitem{Regal2004}
C.~A. {Regal}, M. {Greiner}, and D.~S. {Jin}, Phys. Rev. Lett. {\bf 92},
  040403  (2004); M.~W. {Zwierlein} {\it et~al.}, {\it ibid.} {\bf 92},  120403  (2004).

\bibitem{LiuPRL2005}
X.-J. {Liu}, P.~D. {Drummond}, and H. {Hu}, Phys. Rev. Lett. {\bf 94},  136406
  (2005).

\bibitem{RigolPRL2003}
M. {Rigol}, A. {Muramatsu}, G.~G. {Batrouni}, and R.~T. {Scalettar}, Phys. Rev.
  Lett. {\bf 91},  130403  (2003).

\bibitem{RigolPRB2006}
M. {Rigol}, R.~T. {Scalettar}, P. {Sengupta}, and G.~G. {Batrouni}, \prb {\bf
  73},  121103  (2006).

\bibitem{GeorgesRevMod1996}
A. {Georges}, G. {Kotliar}, W. {Krauth}, and M.~J. {Rozenberg}, Reviews of
  Modern Physics {\bf 68},  13  (1996).

\bibitem{MetznerVollhardt}
W. {Metzner} and D. {Vollhardt}, Phys. Rev. Lett. {\bf 62},  324  (1989).

\bibitem{Dobrosavljevic1997}
V. {Dobrosavljevi{\'c}} and G. {Kotliar}, Phys. Rev. Lett. {\bf 78},  3943
  (1997).

\bibitem{Okamoto2004}
S. {Okamoto} and A.~J. {Millis}, \nat {\bf 428},  630  (2004); \prb {\bf 70},  241104  (2004).

\bibitem{Potthoff}
M. Potthoff and W. Nolting, Phys. Rev. B {\bf 59},  2549  (1999);
S. Schwieger, M. Potthoff, and W. Nolting, {\it ibid.} {\bf 67},  165408
  (2003).

\bibitem{Freericks}
  P. Miller and J.~K. Freericks, J. Phys. Cond. Mat. {\bf 13}, 3187 (2001);
  J.~K. {Freericks} and
  L. {Chen}, Phys. Rev. B {\bf 75},  125114  (2007);
  J.~K. Freericks, {\it Transport in Multilayered Nanostructures}
  (Imperial College Press, London, 2006).

\bibitem{Dobrosavljevic1998}
M.~P. Sarachik {\it et~al.}, Phys. Rev. B {\bf 58},  6692  (1998).

\bibitem{BullaPRB2001}
R. {Bulla}, T.~A. {Costi}, and D. {Vollhardt}, \prb {\bf 64},  045103  (2001).

\bibitem{Bulla2007}
R. Bulla, T. Costi, and T. Pruschke, Rev. Mod. Phys., in press;
  arXiv:cond-mat/0701105  (2007).

\bibitem{Gerbier2007}
F. {Gerbier}, S. {Foelling}, A. {Widera}, and I. {Bloch},
  arXiv:cond-mat/0701420v1  (2007).

\end{thebibliography}
\end{document}